\documentstyle[12pt]{article}

\def\Upp{U^{\prime\prime}}
\def\Up{U^{\prime}}

\def\phibar{{\bar \phi}}
\def\psibar{{\bar \psi}}

\def\ptilde{\tilde p}

\begin{document}

\begin{titlepage}
\noindent January, 1993 \hfill                 BUHEP-93-1\\
\begin{center}
{\bf Does the 2d Higgs-Yukawa Model Have a Symmetric Phase\\
at Small Yukawa Coupling Region?}

\vspace{0.5cm}
Yue Shen\\
Physics Department, Boston University, Boston, MA 02215, USA\\

\vspace{0.5cm}
\end{center}

\abstract{
We show that at arbitrary value of the scalar self coupling and small Yukawa
coupling $y$ the 2d
Higgs-Yukawa model with Z(2) symmetry remains in the broken phase and
the model is asymptotically free: $y \to 0$ as the cut-off $\Lambda \to
\infty$.
This is in agreement with a recent conjecture based on numerical simulation
results.
}
\vfill
\end{titlepage}

The Higgs-Yukawa model has been shown to be equivalent to a generalized
Nambu-Jona-Lasinio (NJL) model in $2< d \le 4$ dimension \cite{UCSD,Zinn}.
However, much less is known for its properties in $d=2$. The lattice action
for the Higgs-Yukawa model with a Z(2) symmetry is given by
\begin{equation}
S = S_h[\phi] +
\sum_{x,z} \psibar_i(x) M(x,z) \psi_i(z)~~, ~~
i = 1,2,...,N_f ~~,
\label{eq:Z2act}
\end{equation}
where the scalar action $S_h[\phi]$ has the form
\begin{equation}
S_h[\phi] = -2\kappa\sum_{x,\mu} \phi(x)\phi(x+\mu)
+ \sum_x \phi^2(x) + \sum_x\lambda \left[\phi^2(x)-1\right]^2 ~.
\end{equation}
The fermion matrix $M(x,z)$ in Eq. (\ref{eq:Z2act}) may be written as
\begin{equation}
M(x,z) = \sum_{\mu} \gamma_\mu\left[\delta_{x+\mu,z}-\delta_{x-\mu,z}\right]
+ y\phi(x)\delta_{x,z} ~,
\label{eq:F-matrix}
\end{equation}
where $\gamma_\mu, \mu=1,...,d$, are the Hermitian Dirac matrices and $y$
stands for the Yukawa coupling. At $\kappa = 0, \lambda = 0$ one can integrate
out the scalar field exactly and obtains a lattice regularized Gross-Neveu (GN)
model (e.g. NJL model for $d=2$). The GN model was solved exactly in
the large $N_f$ limit and it was found that the system always remains in the
broken phase and the theory has asymptotic freedom \cite{Gross}.
On the other hand, at $y= 0$ the fermions will decouple.
The model is believed to be in
the same universality class of 2d Ising model and the system is expected to
have a symmetry breaking phase transition at $\kappa_{c}(\lambda)$. Now the
interesting question is: when $y$ is turned on gradually at $\kappa \ne 0$ and
$ \lambda \ne 0$, is the model
Eq. (\ref{eq:Z2act}) equivalent to some generalized GN model? If so, does it
have a symmetric phase?
In a recent work \cite{De} a numerical simulation was performed for this model
at $\lambda=0$ and $\lambda=0.5$. Combining the numerical results and a
large $N_f$ expansion analysis, which is a good approximation only for small
$\lambda$,
it was conjectured \cite{De} that at arbitrary values of $\lambda$ this model
will remain
in the broken phase at small $y$ and the system is asymptotically free:
$y \to 0$ as the cut-off $\Lambda \to \infty$ for fixed physical quantities.
In this note, we will show that this conjecture is indeed true.

Let us regularize the infrared divergences in 2d by putting the system
in a finite box of linear size $L$. Separating the scalar field into
$\phi(x) = \phibar + \varphi(x)$, where
$\phibar$ is the zero-momentum component and $\varphi(x)$ the nonzero momentum
part with constraint $\sum_x \varphi(x) = 0$, we consider the effetive
potential in a finite volume defined by \cite{Kuti}
\begin{equation}
\exp[-L^d U_L(\phibar)] = \int \prod_x d\varphi(x) \delta\left(\sum_x
\varphi(x)\right) e^{-S_h[\phibar+\varphi]+N_f tr\ln M(\phibar+\varphi)}~,
\end{equation}
where the fermion field has been integrated out and its contribution is
given by $N_f tr \ln M$ term.
For small bare Yukawa coupling $y$ the effects of the fermions can be treated
perturbatively by expanding the fermion determinant in $y^2$. We get
to leading order
\begin{equation}
tr\ln M(\phibar + \varphi) = tr\ln M(\phibar) + O(y^2)~.
\end{equation}
Thus the effective potential is given by \cite{Shen}
\begin{equation}
U_L(\phibar) = U_{hL}(\phibar) - N_f tr\ln M(\phibar) ~,
\label{eq:EFPT}
\end{equation}
where $U_{hL}(\phibar)$ is the {\it exact} effective potential for a
pure scalar system. We emphasize that Eq. (\ref{eq:EFPT}) is only perturbative
in $y$ but nonperturbative in $\lambda$. Although we can not calculate $U_{hL}$
analytically at strong $\lambda$, it can be studied easily in numerical
simulations \cite{Kuti}. The minimum of the effective potential can be obtained
by setting
\begin{equation}
\Up_L(\phibar) = \phibar\left[{\Up_{hL}(\phibar)\over \phibar} - {1\over L^d}
\sum_{p \ne 0} {C_\gamma N_f y^2 \over \ptilde^2 + y^2\phibar^2} \right] = 0~,
\label{eq:VEV}
\end{equation}
where $\ptilde^2 = \sum_\mu sin^2(p_\mu)$ and $C_\gamma$ is the dimension of
the Dirac matrices.
For sufficiently large $\kappa$, the effective potential will have a minimum
at $\phibar = v_L \ne 0$. As long as we approach the critical line from the
broken phase, $v_L$ will have a finite infinite volume limit: $v_L \to v,
L \to \infty$. At the critical
line $v \to 0$, we get the equation for the critical surface
\begin{equation}
\Upp_{hL}(0) - {1\over L^d}
\sum_{p \ne 0} {C_\gamma  y^2 \phibar \over \ptilde^2 } = 0~,
\label{eq:CS}
\end{equation}
where it is understood that the limit $L \to \infty$ is to be taken in the end.
The effect of fermions is to increase the ferromagnetic order so we expect
that $\kappa$ value on the critical surface will satisfy
$\kappa < \kappa_c(\lambda)$, where $\kappa_c(\lambda)$ is the critical point
for the pure scalar model. Consequently $\Upp_{hL}(0)$ is to be evaluated in
the symmetric phase of the scalar model (In numerical simulations one may
measure $<\phibar^2>_L$ to get $\Upp_{hL}(0) = 1/L^d <\phibar^2>_L$).
It has a well defined $L \to \infty$ limit
and we have $\lim_{L \to \infty} \Upp_{hL}(0) = m^2_R/ Z_\phi$ with $m_R$ the
renormalized mass and $Z_\phi$ the wavefunction normalization constant of the
scalar model. So the critical value for $y$ is determined by
\begin{equation}
y^2 = \lim_{L \to \infty} {m_R^2 \over C_\gamma N_f Z_\phi {1\over L^d} \sum_{p
\ne 0}
{1\over \ptilde^2} }~.
\end{equation}
The sum $(1/L^d)\sum_{p \ne 0} 1/\ptilde^2$ is infrared divergent in 2d
\begin{equation}
{1\over L^2}\sum_{p \ne 0} {1\over \ptilde^2} = 4 \cdot {1\over 2\pi}\ln L +
...~,
\end{equation}
where the factor $4$ reflects doubling of the fermions.
So the critical line is at $y \to 0$.

Next we show that Eq. (\ref{eq:VEV}) implies asymptotic freedom.
Let us hold $\kappa < \kappa_c(\lambda), \lambda$ fixed and tune $y$ to
the critical line $y = 0$. For any nonvanishing $y$ Eq. (\ref{eq:VEV}) has
finite $L \to \infty$ limit
\begin{equation}
{\Up_h(v) \over v} - 2N_f y^2 \int {d^2p\over (2\pi)^2} {1\over \ptilde^2
+ y^2v^2} =0~.
\label{eq:Gap}
\end{equation}
For small $y v$ we have \cite{Kogut}
\begin{equation}
\int {d^2p\over (2\pi)^2} {1\over \ptilde^2
+ y^2v^2} = 4\left\{-{1\over 2\pi} \ln (yv) + {1\over 4\pi}\ln 8
+O\left(y^2v^2\ln(yv)\right)
\right\}~,
\end{equation}
where again the factor $4$ comes from fermion doubling.
In the symmetric phase, the scalar potential has the form
\begin{equation}
\Up_h(v)/v = {m_R^2 \over Z_\phi} + c_4 v^2 + c_6 v^4 + O(v^4) ~,
\end{equation}
where $c_4, c_6, ...$ are functions of $\kappa$ and $\lambda$. So we obtain
from Eq. (\ref{eq:Gap})
\begin{equation}
y v = 2\sqrt{2} \exp \left\{-{\pi \over 4N_f y^2}{m_R^2 \over Z_\phi} +
O\left({1\over y^4}
e^{-{const \over y^2}} \right) \right\}~,
\end{equation}
where the neglected terms are exponentially small in $1/y^2$.
The fermion propagator can be calculated also in the leading $y$ expansion
\begin{equation}
G_f(p) = <\psi(p)\psibar(-p)> = {1 \over i\sum_\mu \gamma_\mu sin(p_\mu)
+ yv}~.
\end{equation}
In the continuum limit, the fermion mass is given by the pole of $G_f(p)$
and we have
\begin{equation}
m_f = {1\over a}yv =  {1\over a} 2\sqrt{2} \exp\left\{-{\pi \over 4N_f y^2}
{m_R^2\over Z_\phi}\right\}~,
\end{equation}
where the explicit cut-off $\Lambda = 1/a$ dependence is restored.
When the cut-off is removed
$a \to 0$, in order to keep the fermion mass fixed we have to tune $y \to 0$
and this is the well known asymptotic freedom.

The generalization of this method to the continuous chiral symmetric
models in 2d is tricky due to the fact that there is no finite vacuum
expectation value in the infinite volume limit.
But the idea that the fermion part can be
treated as a perturbative correction to the nonperturbative scalar sector
might still be valid.
In contrast, application of this method to a continuous chiral symmetric
model in $2 < d \le 4 $ is straight forward. For example, the critical surface
of a $U(1)$ chiral symmetric model \cite{Anna,UCSD,HJS} in 4d is given by
\begin{equation}
{m_R^2 \over Z_\phi} - 4N_fy^2 \int {d^4p \over (2\pi)^4} {1\over
\ptilde^2}=0~,
\label{eq:SU2}
\end{equation}
with
$$
\int {d^4p \over (2\pi)^4} {1\over \ptilde^2}= 16 \cdot 0.03873...~~.
$$
Because only the amplitude of the zero mode $\phibar$ appears in the fermion
determinant this same formula also describes the critical surface of
a $SU(2) \times SU(2)$ symmetric model \cite{Bock}.
As shown in Fig. 1, direct comparison with the $SU(2)\times SU(2)$ model
simulation results of ref \cite{Bock} shows very good agreement.
The difference between 2d and 4d is obvious: because the integral
$\int {d^dp \over (2\pi)^d} {1\over \ptilde^2}$ is infrared finite in $d>2$
there is a finite symmetric region at small $y$ in 4d while this symmetric
region is shrank to $y=0$ axis in 2d due to the infrared divergence.

We should emphasize that the method used here is only valid in the small Yukawa
coupling region. Indeed, in the strong Yukawa coupling region, using either a
large $y$ expansion or a large $N_f$ expansion with
$y^2 \sim N_f$ \cite{Affl,UCSD,HJS} one may show that this model
is equivalent to the 2d Ising model and thus has a symmetric phase.

In conclusion, we have shown that the 2d Higgs-Yukawa model with Z(2)
symmetry has no symmetric phase in the small $y$ region. The vacuum expectation
value $v$ satisfies
the gap equation Eq. (\ref{eq:Gap}) and the system is asymptotically free.
All these properties suggest that the continuum limit of this model
is in the same universality class of the GN model.

\noindent{\bf Acknowledgement}

I am grateful to J. Kuti for encouragement and thank him, R.~Brower and
K.~Jansen for discussions.
This work was supported in part under DOE contract DE-FG02-91ER40676 and NSF
contract PHY-9057173, and by funds from the Texas National Research Laboratory
Commission under grant RGFY92B6.

\pagebreak

\noindent{\bf Figure 1}: The critical line separating the ferromagnetic (FM)
phase and the symmetric (SYM) phase at $\lambda=\infty$ for
$SU(2)\times SU(2)$ symmetric model in 4d. The solid line is predicted by
Eq. \protect{(\ref{eq:SU2})}
with $N_f = 2$. $m_R^2/Z_\phi$ is measured in a simulation of the $O(4)$
symmetric scalar model. The data points are taken from
\protect {Ref \cite{Bock}}. Note that in 2d the symmetric region will disappear
and the critical line will shift to $y \to 0$.


\begin{thebibliography}{99}

\bibitem{UCSD}A.~Hasenfratz, P.~Hasenfratz, K.~Jansen, J.~Kuti
and Y.~Shen, Nucl. Phys. B365 (1991) 79.

\bibitem{Zinn}J.~Zinn-Justin, Nucl. Phys. B367 (1991) 105.

\bibitem{Gross}D.~J.~Gross and A.~Neveu, Phys. Rev. D10 (1974) 3235.

\bibitem{De}A.~De, E.~Focht, W.~Franzki and J.~Jers{\' a}k, preprint
HLRZ-92-97,
hep-lat/9212024.

\bibitem{Kuti}J.~Kuti and Y.~Shen, Phys. Rev. Lett. 60 (1988) 85.

\bibitem{Shen}Y.~Shen, J.~Kuti, L.~Lin and P.~Rossi,
Nucl. Phys. B (Proc. Suppl.) 9 (1989) 99.

\bibitem{Kogut}J.~Kogut and J.~Shigemitsu, Nucl. Phys. B190 (1981) 365.

\bibitem{Anna}A.~Hasenfratz, W.~Liu and T.~Neuhaus,
Phys. Lett. B236 (1990) 339.

\bibitem{HJS}A.~Hasenfratz, K.~Jansen and Y.~Shen, BNL preprint, BNL-47708.

\bibitem{Bock}W.~Bock, A.~K.~De, K.~Jansen, J.~Jers{\' a}k, T.~Neuhaus
and J.~Smit, Nucl. Phys. B344 (1990) 207.

\bibitem{Affl}I.~K.~Affleck, Phys. Lett. B109 (1982) 307.

\end{thebibliography}
\end{document}